\documentclass[12pt,preprint]{aastex}

\newcommand{\hhh}{H$_3^+$}

\begin{document}

\title{\hhh\ in Diffuse Interstellar Clouds: A Tracer for the Cosmic-Ray
Ionization Rate}

\author{Nick Indriolo\altaffilmark{1},
Thomas R. Geballe\altaffilmark{2},
Takeshi Oka\altaffilmark{3},
and Benjamin J. McCall\altaffilmark{1}}

\altaffiltext{1}{Department of Astronomy and Department of Chemistry, University of
Illinois at Urbana-Champaign, Urbana, IL 61801; nindrio2@uiuc.edu, bjmccall@uiuc.edu}
\altaffiltext{2}{Gemini Observatory, 670 North A'ohoku Place, Hilo, HI 96720}
\altaffiltext{3}{Department of Astronomy \& Astrophysics and Department of 
Chemistry, University of Chicago, Chicago, IL 60637}

\begin{abstract}
Using high resolution infrared spectroscopy we have surveyed twenty sightlines for \hhh\   
absorption.  \hhh\ is detected in eight diffuse cloud
sightlines with column densities varying from $0.6\times10^{14}$~cm$^{-2}$ to 
$3.9\times10^{14}$~cm$^{-2}$. 
This brings to fourteen the total number of diffuse cloud sightlines where \hhh\ has been detected.
These detections are mostly along sightlines concentrated in the Galactic plane, but well dispersed 
in Galactic longitude.  The results imply that abundant \hhh\ is common
in the diffuse interstellar medium.  Because of the simple chemistry associated with \hhh\ production and destruction,
these column density measurements can be used in concert with various other data to infer the primary cosmic-ray ionization
rate, $\zeta_p$.  Values range from  $0.5\times10^{-16}$~s$^{-1}$ to $3\times10^{-16}$~s$^{-1}$ with an average of 
$2\times10^{-16}$~s$^{-1}$. Where \hhh\ is not detected the upper limits on the ionization rate are consistent with this range.
The average value of $\zeta_p$ is about an order of magnitude larger than both the canonical rate and rates previously reported
by other groups using measurements of OH and HD. The discrepancy is most likely due to inaccurate measurements of rate constants
and the omission of effects which were unknown when those studies were performed.  
We believe that the observed column density of \hhh\ is the most direct tracer for the cosmic-ray ionization rate due to its simple
chemistry.  Recent models of diffuse cloud chemistry
require cosmic-ray ionization rates on the order of 10$^{-16}$~s$^{-1}$ to reproduce observed abundances of various  atomic and
molecular species, in rough accord with our observational findings.
\end{abstract}


\section{INTRODUCTION}

In the past several years \hhh\ has been detected in diffuse interstellar clouds \citep{mcc98,mcc02,geb99} where it had been
expected to exist in abundances below observable limits.  This surprising result raised various questions about the diffuse
cloud environment.  The rather simple chemistry of \hhh\ allows for only three variable parameters in determining its abundance
when the steady state approximation is used: the \hhh-electron recombination rate, the electron to
hydrogen ratio, and the cosmic-ray ionization rate.  Previous work \citep{mcc03,mcc04,car96} has shown that the first two of
these are relatively well constrained.  This leaves the cosmic-ray ionization rate as an unconstrained parameter.
Because the low energy cosmic-rays responsible for most of the ionization in diffuse clouds cannot be directly 
measured in the solar system, we must rely on molecules to act as tracers of the ionization rate.  
Using \hhh, \citet{mcc03} found the cosmic-ray ionization rate of molecular hydrogen, $\zeta_2$, to be much larger
along the sightline to $\zeta$~Per than the canonical value of $\sim 3\times10^{-17}$~s$^{-1}$.

Prior to the detection of \hhh\ in diffuse clouds, OH and HD were the molecules of choice for estimating
the cosmic-ray ionization rate there.  Estimates using these molecules required determining 
rate constants and modeling various reactions on
the pathways to forming OH and HD \citep{bla77,fed96,odo74}. The derived values of the ionization rate tended to agree with
the canonical value of $\zeta_p$, the primary cosmic-ray ionization rate, but differ greatly from the value derived from  
the recent \hhh\ measurement toward $\zeta$~Per (the relation between $\zeta_p$ and $\zeta_2$ is explained in \S~4.2 and quantified
by equation (\ref{eq10})).  
Of the three molecules, the simple chemistry of \hhh\ provides the most direct determination of $\zeta_p$
\citep{dal06}, suggesting that measurements of \hhh\ should produce more accurate results and be a more reliable tracer of the
cosmic-ray ionization rate than OH or HD.

The higher ionization rate found by \citet{mcc03} towards $\zeta$ Per implies the production of more \hhh, 
and if generally applicable, could
account for the higher than expected column densities found in several diffuse clouds \citep{mcc02}. However, prior to
the present work the enhanced ionization rate was known to exist for certain only along one line of sight, 
and thus could have been considered an anomaly.  
To test if an enhanced ionization rate is a general property of the diffuse interstellar medium (ISM), we have performed 
a survey of \hhh\ in nineteen diffuse
cloud sightlines.  \hhh\ is detected in eight of the clouds and the overall results, including analysis of previous
observations by our group, support a higher ionization rate.  When coupled with further arguments, this strongly suggests 
that a greatly enhanced ionization rate is a typical property of the diffuse ISM.

\section{METHODS}

\subsection{Observations}

All observations were made using the CGS4 spectrometer \citep{mou90} on the United Kingdom Infrared Telescope (UKIRT)
at Mauna Kea.  The spectrometer was used with its echelle grating, 0.6 arcsec wide slit, and long camera to provide a resolving
power of 40000.  Observations were taken in an ABBA pattern in which the target star is nodded along the slit so that the
spectral image falls alternately on different rows of the array.  Suitable standard stars were observed
throughout each night to account for changing atmospheric conditions and air mass.  With the exception of the 
Red Rectangle where reddening is due to the ejected envelope of a post asymptotic giant branch star, targets 
were chosen primarily by
three criteria: (1) sightlines known to pass through diffuse clouds; (2) early spectral type; and (3) bright L-band
magnitude. The complete dataset consists of twenty (nineteen diffuse cloud) sightlines which were observed 
intermittently between May 2001 and March
2005.  Relevant information concerning these observations is given in Table \ref{tbl1}. Two of the sightlines from the dataset
were examined in \citet{mcc02}: HD~20041 and $\zeta$~Oph. However, both have previously been studied using only the $R(1,1)^l$
transition at 37154.8 \AA\ (vacuum wavelengths are used throughout this paper), whereas the new data cover the 
$R(1,1)^u$ and $R(1,0)$ transitions at 36680.8~\AA\ and 36685.2~\AA,
respectively (see \citet{mcc00} or \citet{mcok00} for a complete description of the 
transition notation associated with \hhh).
$\zeta$~Per was investigated in \citet{mcc03}, but here data from three more nights
of observations are included. Overall, we
present twenty sightlines for which new or refined \hhh\ column densities or upper limits are calculated.

\subsection{Data Reduction}
The reduction process involves multiple steps and software packages.  First, raw data are run through Starlink's 
ORAC-DR\footnote{http://www.oracdr.org/} pipeline which processes UKIRT data.  Images are then transformed to fits format
using Starlink's FIGARO\footnote{http://www.starlink.rl.ac.uk/star/docs/sun86.htx/sun86.html} package, and neighboring images
are subtracted from each other in NOAO's 
IRAF\footnote{http://iraf.noao.edu/} package.  This subtraction serves to eliminate atmospheric background and detector bias
levels from the image.  Still using IRAF, spectra are extracted from the neighbor-subtracted images with the {\it apall}
routine.  These spectra are imported to IGOR Pro\footnote{http://www.wavemetrics.com/} where we have macros set up to complete
the reduction \citep{mccth}.  
During this import, ripples in the spectrum caused by guiding and seeing fluctuations while the CGS4 array is shifted along the
spectral direction in successive steps of one-third of a pixel during observing are removed. 
Spectra for each object and standard star are then coadded.  Some frames
may be excluded in this step if they are noisier than normal or happen to have cosmic-ray hits on the rows where the
spectrum is located. Objects are then ratioed with standards to remove atmospheric absorption lines and the continuum level is
set to unity. The ratioing is an interactive process where the user may vary the intensity scaling and/or shift the spectrum
in wavelength in order to most effectively
remove atmospheric lines and obtain the most reliable ratioed spectrum at the wavelengths where the \hhh\ lines are expected
to appear.  A fringing pattern caused by the circular variable order-blocking filter is another artifact of CGS4 that
needs to be removed.  This is accomplished by transforming a spectrum into Fourier space using an IGOR macro.  The
user can then find the peak caused by the fringing pattern and interpolate across it.  Once the peak is removed, an inverse
Fourier transform is performed to produce a spectrum where the fringing pattern is absent.  This method is described in more
depth in \citet{mccth}. After removing the fringing pattern, spectra are wavelength calibrated using the vacuum rest 
wavelengths of the atmospheric lines.  The accuracy of the wavelength calibration is typically $\pm$2~km~s$^{-1}$.  
Finally, the \hhh\
lines are fit with Gaussians and the equivalent widths, column densities, and radial velocities are derived.  

For targets with observations on multiple nights, the reduction process above is followed
through the wavelength calibration step.  At that point IRAF's {\it rvcorrect} routine is used to calculate the velocity of
the Earth along a given sightline in the local standard of rest (LSR) frame.  Each spectrum is then shifted to be in the LSR
frame. This puts the interstellar absorption lines at the same wavelength for any given date and allows spectra from different
nights to be coadded.  While each spectrum has its continuum set to unity, they originally had different exposure times and
intensities.  To weight the final coadded spectrum properly, each spectrum is scaled by its original coadded
intensity before being added to the spectra from other nights.  Once the final multiple night coadded spectrum is produced, it
is divided by the sum of the scaling factors to reset the continuum level to unity.  After this process, line parameters are
again extracted by fitting the absorption lines.

\subsection{Atmospheric Interference}

\begin{figure}
\epsscale{.8}
\plotone{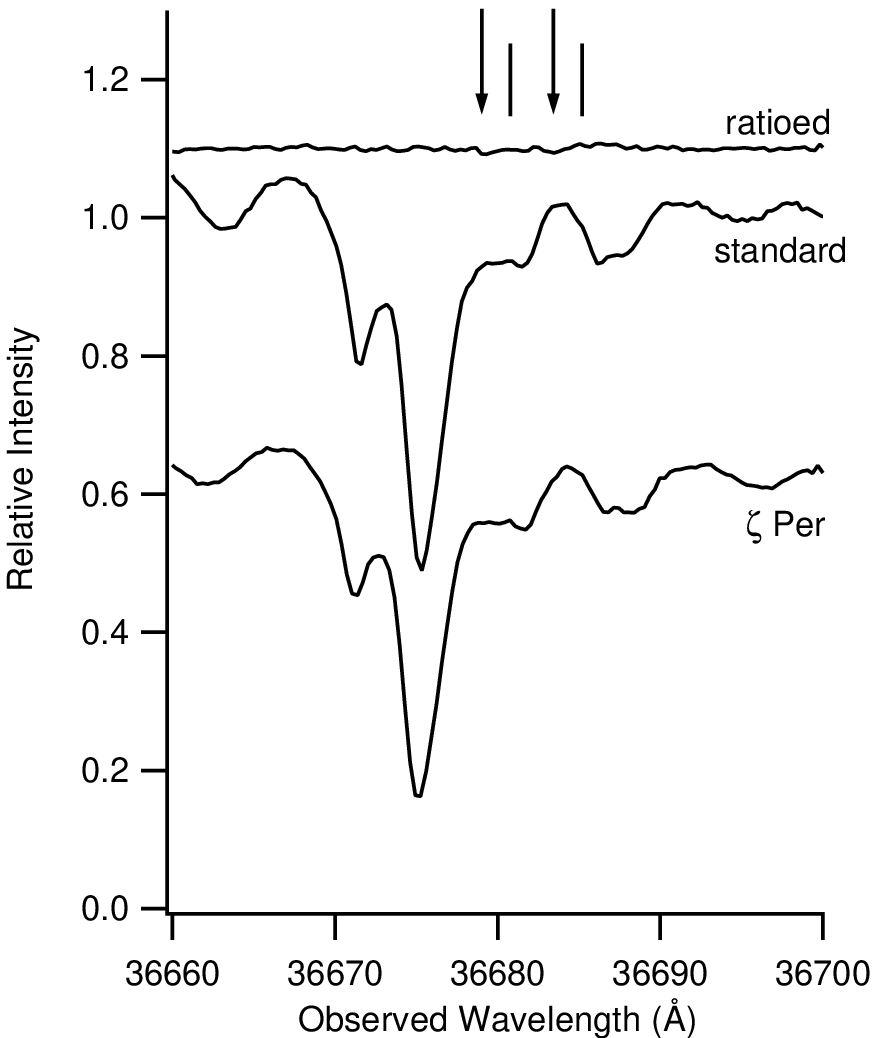}
\caption{Typical examples of spectra near the \hhh\ ortho-para doublet.  The spectra in this and all other figures
have been offset in intensity for clarity. 
The bottom spectrum is $\zeta$~Per from 2001 September 5.
The middle spectrum is the standard star 
$\beta$~Per from the same date.  The top spectrum is $\zeta$~Per ratioed with the standard star.  
Two arrows show the expected
location of \hhh\ absorption which can barely be seen here due to the scaling.  The vertical lines are at the rest 
wavelengths of the \hhh\ lines.}
\label{fig1}
\end{figure}

Spectra covering the \hhh\ doublet are adversely affected by an atmospheric CH$_4$ line complex which lies just shortward of 
36680.8 \AA, the rest wavelength of the $R(1,1)^u$ line.  If the spectrum of the target is blue-shifted, the overlap with this
line can be significant and hinder detection or estimates of line strength. Typical examples of spectra before the atmospheric
lines are removed via ratioing with a standard star are shown in Figure \ref{fig1}.  The CH$_4$ feature here absorbs roughly
50\% of the incoming light at 36675.3~\AA\ in both the standard star $\beta$~Per (middle spectrum) and the object
$\zeta$~Per (bottom spectrum).  A weak telluric HDO absorption line near 36681~\AA\ that can vary with both time and
airmass further complicates the reduction, especially when the water column density above the telescope is high and unstable.
When the two spectra are ratioed following the methods in \S~2.2, the top spectrum in Figure \ref{fig1} is
produced.  The two lines show the rest wavelength positions of the $R(1,1)^u$ line at 36680.8~\AA\ and the $R(1,0)$ line at 
36685.2~\AA.  Two arrows mark the expected positions of the \hhh\ lines due the Earth's orbital motion and the radial
velocity of the absorbing cloud along the line of sight towards $\zeta$~Per.  Clearly the \hhh\ absorption lines are much
weaker than the atmospheric absorption lines and are barely visible at this scaling factor.  This illustrates why a careful
multi-step reduction process is necessary to detect \hhh.

\section{RESULTS}
\subsection{Positive Detections}

\begin{figure}
\epsscale{.8}
\plotone{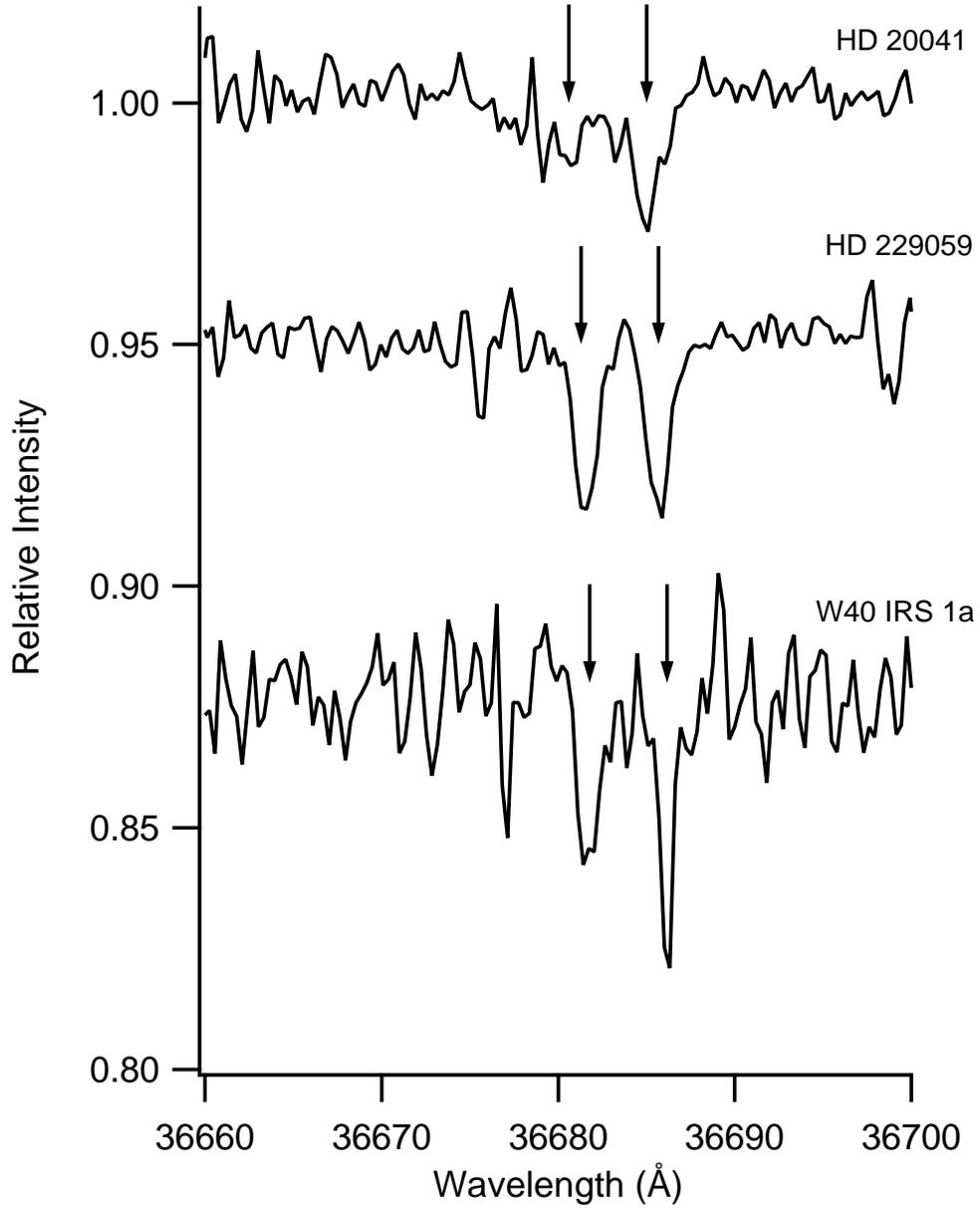}
\caption{Spectra showing strong detections of the \hhh\ doublet near 36680~\AA . 
All spectra have been Doppler shifted into the rest frame of the LSR.         
Arrows show where the lines are expected due to previous gas velocity
measurements, which are given in Table \ref{tbl2}.}
\label{fig2}
\end{figure}

\begin{figure}
\epsscale{.8}
\plotone{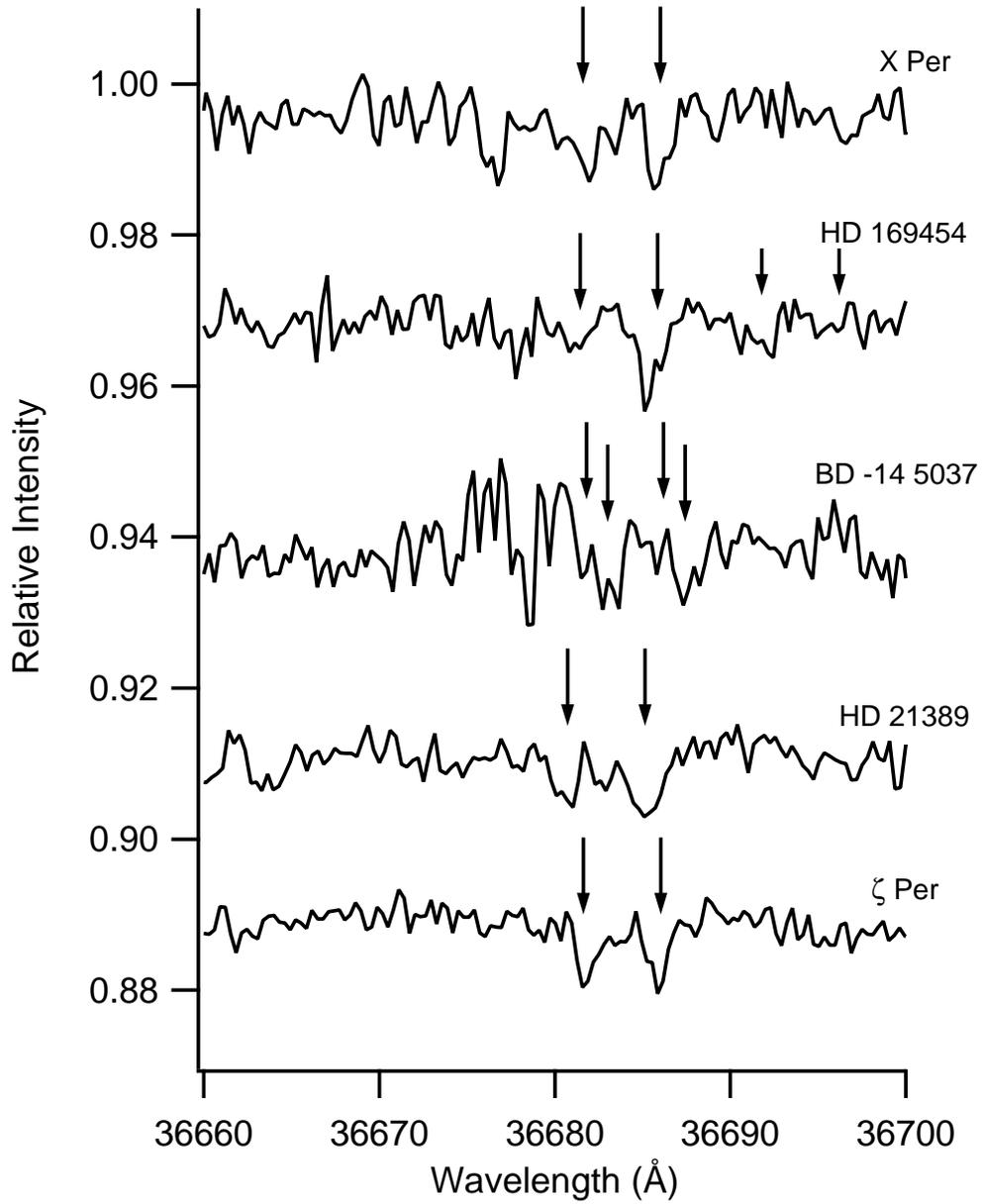}
\caption{Same as Figure 2 except showing more typical strength detections of \hhh.}
\label{fig3}
\end{figure}

\begin{figure}
\epsscale{.8}
\plotone{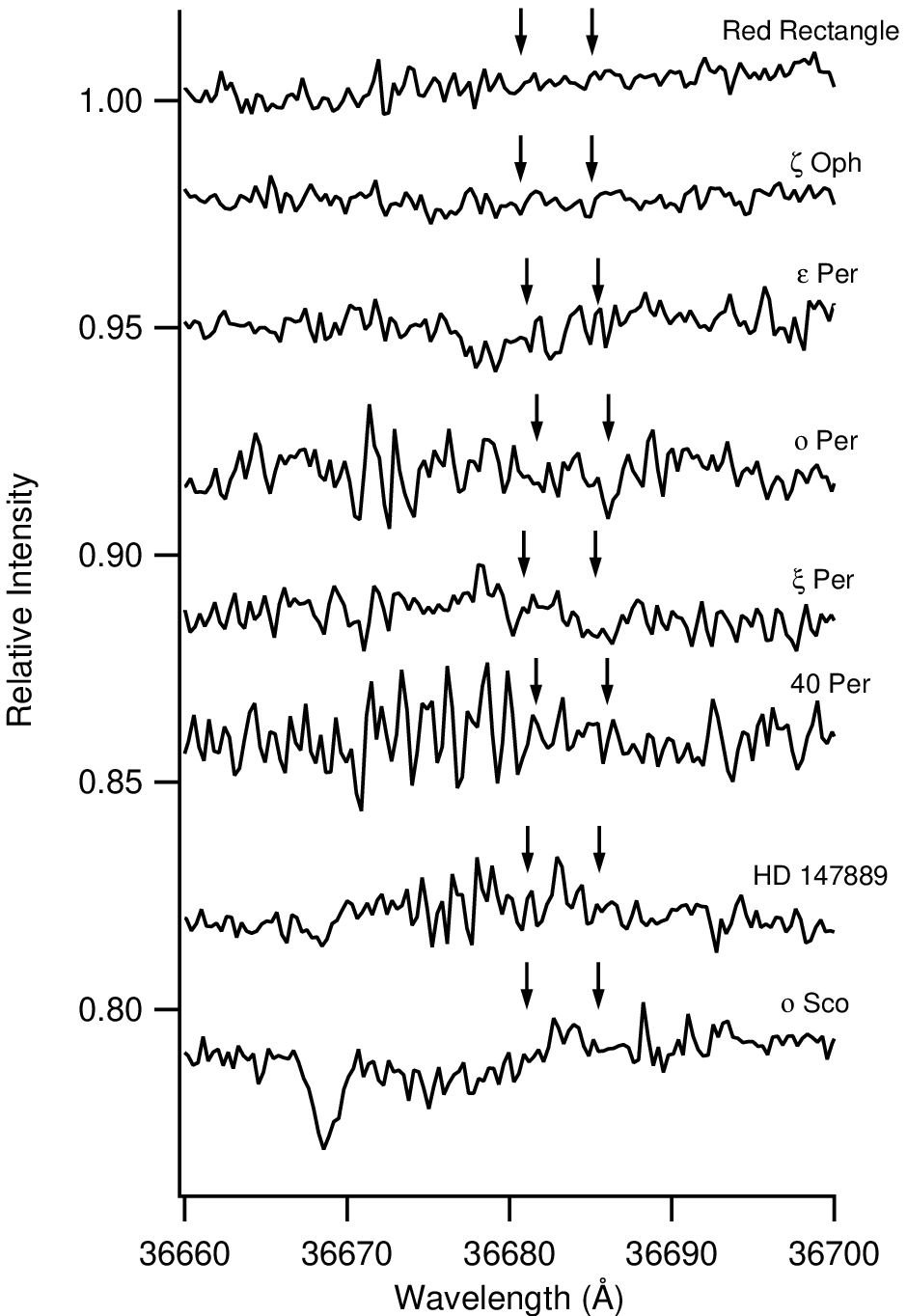}
\caption{Same as Figure 2 except showing non-detections.}
\label{fig4}
\end{figure}

\begin{figure}
\epsscale{.8}
\plotone{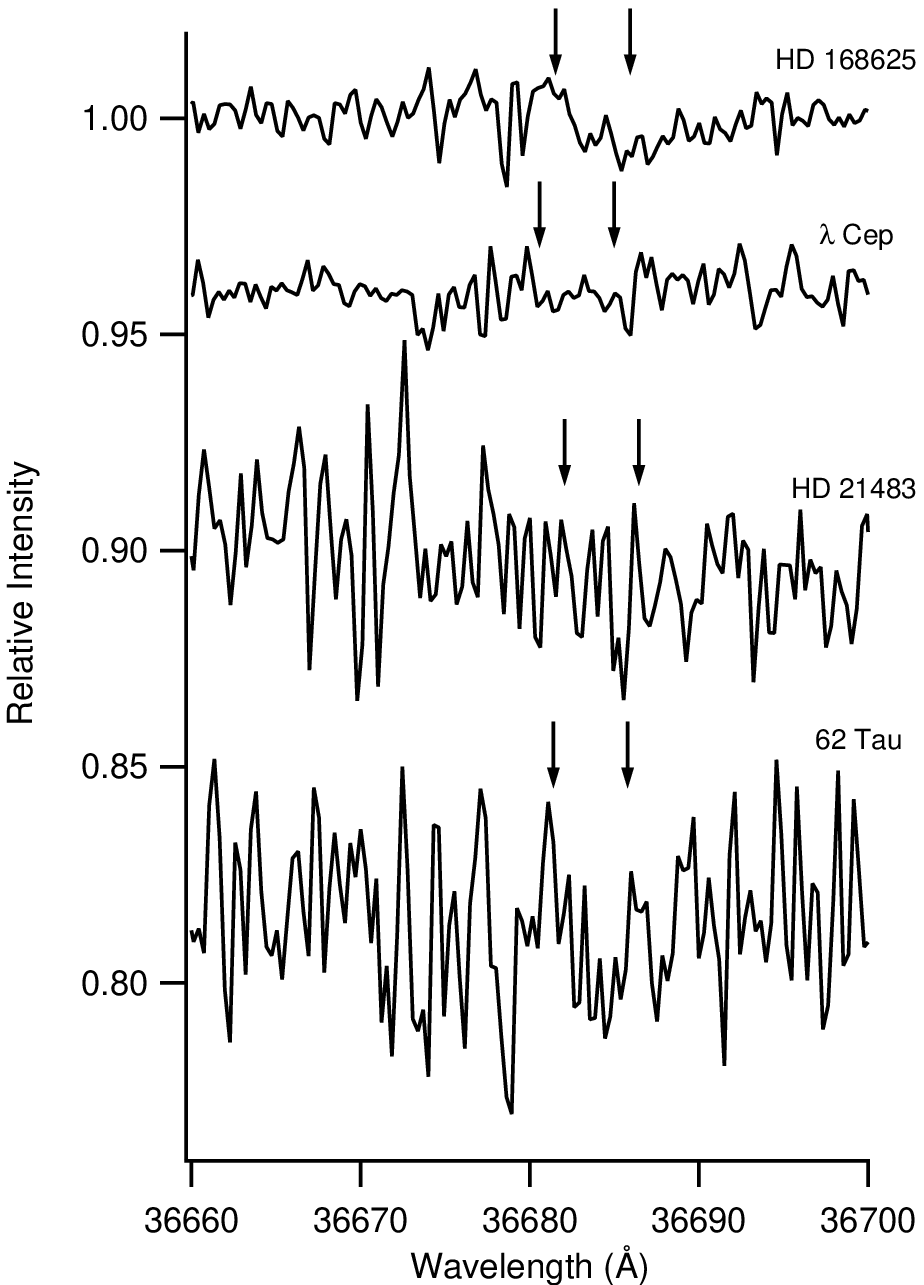}
\caption{Same as Figure 2 except showing non-detections.}
\label{fig5}
\end{figure}

The fully reduced spectra are shown in Figures \ref{fig2}-\ref{fig5}.  Figures \ref{fig2} and \ref{fig3} contain 
spectra from  sightlines with positive \hhh\ detections, and Figures \ref{fig4} 
and \ref{fig5} show spectra from sightlines with no \hhh\ detections or marginal detections.  Arrows indicate the
position of the \hhh\ doublet expected from previous measurements of the gas 
velocity along each line of sight.  These velocities are
given in Table \ref{tbl2} along with the atomic or molecular species from which they were determined.  In the figures the
typical noise level can be judged by the peak-to-peak fluctuations in the continuum well off of the \hhh\  lines.  Within several
\AA ngstr\"{o}ms of 36675.3~\AA, however, the noise in the reduced spectra is several times larger than elsewhere due to the
presence of the strong complex of CH$_4$ lines, which both reduces the atmospheric transmission and emits excess background.

The three spectra in Figure \ref{fig2} have the strongest \hhh\ absorption.  In each case the lines match up with
the arrows so we are confident in the \hhh\ detection.  In HD~20041 the $R(1,0)$ transition is clearly visible but the 
$R(1,1)^u$ transition may have been affected by the
aforementioned CH$_4$ line.  Both absorption lines are clear and strong in HD~229059.  W40~IRS~1a presents a somewhat confusing
case because there may be either one or two velocity components.  If there is one component, then the two absorption lines
to longer wavelengths represent the \hhh\ doublet and the shorter wavelength line at 36677~\AA\ is a noise artifact.  
If there are
two absorption components, then the $R(1,0)$ transition from the shorter wavelength doublet overlaps the $R(1,1)^u$ transition
from the longer wavelength doublet.  However, the W40~IRS~1a spectrum is composed of data from two nights and on only one of
those nights does the 36677~\AA\ feature appear.  For this reason we conclude that only the longer wavelength doublet is
real. Our conclusion is consistent with the observations reported by \citet{cru82} in which only one velocity component
at about 8~km~s$^{-1}$ LSR was seen in $^{13}$CO emission, but does not match the 2$\pm$2~km~s$^{-1}$ LSR reported by
\citet{shu99} from $^{12}$CO absorption.  Note, however, that if the other line is real then it would correspond to a
cloud with a radial velocity of about -30~km~s$^{-1}$ LSR, which disagrees badly with both CO measurements.

In Figure \ref{fig3} the absorption lines are again aligned with the arrows marking previously observed gas velocities. 
$\zeta$~Per and HD~21389 have the highest signal to noise ratios (SNR) and their \hhh\ lines are easily
identified.  The doublet in X~Per is relatively clear, but the velocities found by fitting the individual line profiles
differ by 2.5~km~s$^{-1}$.  Most likely this is due to noise affecting the absorption feature.  In HD~169454
the $R(1,0)$ transition of the main doublet is clear, but the $R(1,1)^u$ line is rather shallow.  As in the case of HD~20041
above, this may be caused by interference from the telluric CH$_4$ line.  However, the velocity we derive for the $R(1,0)$
transition differs from the previously measured cloud velocity by about 3.2~km~s$^{-1}$.  Because of the different velocities
and the lack of a clear $R(1,1)^u$ line, we are not as confident as in the previous cases that these features represent
\hhh\ absorption, but still consider it a positive detection. The shorter arrows above the HD~169454 spectrum mark the
expected location of the \hhh\ doublet for a high velocity component at 90~km~s$^{-1}$ reported by \citet{fed92}.  Our
spectrum may indicate absorption features at this velocity, but further integration time is needed to determine if the
features are real.  The absorption features in BD~-14~5037 both appear to be double peaked, although the signal-to-noise
ratio of each double peak is low.  It also seems that there are small absorption features on the shorter wavelength shoulders of
the main features.  Two velocity components reported by \citet{gre86} match well with the centers of the double peaked features
and the shoulder features, making the detections more believable.

\subsection{Negative Detections}

There is no definitive evidence for \hhh\ absorption lines in any spectrum in Figure \ref{fig4}, with the possible exception
of $o$~Per.  The arrow marking the $R(1,0)$ transition of \hhh\ in $o$~Per matches rather well with a statistically
significant absorption feature in the spectrum.  However, at the expected wavelength of the $R(1,1)^u$ line the absorption is too
weak to attempt a fit given the noise level.  Both $\xi$~Per and $\epsilon$~Per show some amount of absorption near the 
wavelengths  where the \hhh\ lines are expected, but nothing that can be conclusively identified as a detection. The Red
Rectangle and $\zeta$~Oph sightlines both have a very high SNR.  With these clean spectra and no \hhh\ absorption, it is
possible to derive strict upper limits.  HD~147889, 40~Per, and $o$~Sco all have no significant absorption features close
to the expected wavelengths. Figure \ref{fig5} contains two non-detections in spectra with typical  noise levels
(HD~168625 and $\lambda$~Cep) and two with high noise levels (HD~21483 and 62~Tau).

\section{ANALYSIS}

By fitting the absorption lines in the spectra with Gaussians we are able to obtain the line of sight velocity,
full width at half maximum, equivalent width, and the \hhh\ column density and its uncertainty.  Values for these parameters
along all of the observed sightlines are given in Table \ref{tbl3}.  Using these values in concert with the steady state
approximation and a few reasonable assumptions allows for the calculation of other physical parameters of the diffuse
clouds along these lines of sight.

\subsection{Reactions}

Below are the three reactions which describe the dominant creation and destruction processes for \hhh.  Reactions (\ref{eq1})
and (\ref{eq2}) show the formation process, while reaction (\ref{eq3}) 
shows destruction.                  
\begin{equation}
{\rm CR} + {\rm H}_2 \rightarrow {\rm CR} + {\rm H}_2^+ + {\rm e}^-
\label{eq1}
\end{equation}
\begin{equation}
{\rm H}_2 + {\rm H}_2^+ \rightarrow {\rm H}_3^+ + {\rm H}
\label{eq2}
\end{equation}
\begin{equation}
{\rm H}_3^+ + {\rm e}^- \rightarrow {\rm H}_2 + {\rm H}\ {\rm or}\ 3{\rm H}
\label{eq3}
\end{equation}

First, H$_2$ is ionized to produce H$_2^+$ and an electron.  This ionization is assumed to be due to a
cosmic-ray because \citet{gla74} showed that low energy cosmic-rays will penetrate diffuse clouds while 
the X-ray flux is attenuated in a thin layer at the cloud exterior.  They find that for an X-ray with energy 
100~eV the optical depth of the cloud reaches unity at an H$_2$ column density of 
$2\times10^{19}$ cm$^{-2}$.  Because most of our sightlines have column densities much larger than this 
we assume that cosmic-rays are the only ionization mechanism operating throughout the majority of 
the cloud.  After being
ionized the H$_2^+$ ion reacts with  H$_2$  to produce \hhh\ and H.  The second step is many orders of 
magnitude faster than the first step \citep{mcc98}, so the formation rate of \hhh\ is proportional to the 
product of the ionization rate and H$_2$ density.  In diffuse clouds the primary channel of \hhh\ destruction 
is electron recombination, which results in either three H atoms or one H atom and one H$_2$ molecule.  The 
destruction rate is given by a rate constant times the product of the number densities of \hhh\ and 
electrons.

\subsection{Calculations}

The steady state approximation assumes that the formation and destruction rates
of \hhh\ are equal.  This approximation yields the equation \citep{geb99}
\begin{equation}
n({\rm H}_2)\zeta_2  = k_e n({\rm H}_3^+)n(e)
\label{eq4}
\end{equation}
where $n$(X) is the number density of species X, $\zeta_2$ is the ionization rate of
H$_2$, and $k_e$ is the \hhh-electron recombination rate constant.
In contrast to the steady state approximation,
time dependent models developed by \citet{lis07} showed that the abundance of \hhh\ is only weakly
dependent on the cosmic-ray ionization rate when a cloud is young.  This age is quantified by the
ratio of molecular hydrogen to atomic hydrogen 
$n({\rm H}_2)/n({\rm H})$ where smaller values correspond to younger clouds.
The \hhh\ abundance becomes weakly dependent on the cosmic-ray ionization rate when 
$n({\rm H}_2)/n({\rm H})\leq0.05$.  This value corresponds to a molecular
hydrogen fraction (defined below in equation (\ref{eq6})) of $f\leq0.09$.  Because the H$_2$
fractions in all of the clouds we observed are more than double this value, we 
neglect time dependence and use the steady state approximation.

Assuming that gas is uniformly distributed in the cloud, we can substitute the
column density divided by the path length for the number density.  By doing this
and solving for the ionization rate we obtain
\begin{equation}
\zeta_2=N({\rm H}_3^+)\frac{k_e}{L}\frac{n(e)}{n({\rm H}_2)}.
\label{eq5}
\end{equation}
We further assume that nearly all
hydrogen is either in the atomic or molecular state, and define the
molecular hydrogen fraction (the fraction of hydrogen nuclei in molecular form) as
\begin{equation}
f\equiv\frac{2n({\rm H}_2)}{n({\rm H})+2n({\rm H}_2)}
\label{eq6}
\end{equation}
where the denominator is the number density of hydrogen nuclei, $n_{\rm H}$.  Solving
for $n({\rm H}_2)$ and plugging the result back into the ionization rate
equation we find  
\begin{equation}
\zeta_2=N({\rm H}_3^+)\frac{k_e}{L}\frac{2}{f}\frac{n(e)}{n_{\rm H}}.
\label{eq7}
\end{equation}
In this form it is possible to measure or estimate all of the variables on the
right hand side of the equation, so we can derive values for the 
ionization rate of molecular hydrogen.

The electron recombination rate constant is given by the equation
\begin{equation}
k_e = -1.3\times10^{-8} + 1.27\times10^{-6}T_e^{-0.48}\ ({\rm cm}^3\ {\rm s}^{-1})
\label{eq8}
\end{equation}
from \citet{mcc04} which is valid when the electron temperature, $T_e$, is between 10 K and 4000
K.  While $T_e$ is not directly
measured, it can be approximated by the excitation temperature derived from the $J=0$ and $J=1$
levels of molecular hydrogen, $T_{01}$.  This temperature is calculated from measurements of
the column densities of the two levels.  In sightlines without these measurements we adopt a 
value of 60~K.
The fact that the observed lines of the $J=0$ and $J=1$ levels of H$_2$ are saturated indicates
that few photons are present in the interior of diffuse clouds to radiatively pump these levels.
This means that collisions will dominate the equilibrium between these levels.  However, the   
$J=0$ and $J=1$ levels of H$_2$ have different nuclear spin configurations and thus require
collisions with species such as H$^+$ or \hhh\ to interconvert \citep{sno06}.  If
collisions with protons are the dominant factor in determining the relative population of the 
$J=0$ and $J=1$ levels though, then $T_{01}$ should represent the proton kinetic temperature
and thus the kinetic temperature of the gas in general \citep{sav77}.  While electrons produced
by photoionization may begin with much higher temperatures, they should thermalize quickly via
collisions with H$_2$ \citep{mcc02}.  Because $T_e$ and $T_{01}$ should both nearly equal the
kinetic temperature of the gas, we substitute $T_{01}$ for $T_e$ in equation (\ref{eq8}). 

Assuming that nearly all electrons in diffuse clouds are
produced via the ionization of C to C$^+$
and that nearly all atomic carbon has been singly ionized
\citep{van86}, the carbon to hydrogen ratio
should approximate the electron to hydrogen ratio.  
\citet{car96} found this value to be about $1.4\times10^{-4}$ in multiple diffuse clouds.  Because
of the relative uniformity of this ratio in all six of their sightlines, 
we adopt a single average value for use in all of our calculations.  

The molecular hydrogen fraction is dependent on H and H$_2$ number densities, quantities which surely
vary through the cloud, but whose variations are not readily measurable.
Because fluctuations in number density cannot be directly measured, we use column densities in place of
the number densities in equation (\ref{eq6}) and calculate what \citet{sno06} refer to as $f^N$
in clouds where we
have measurements of the H and H$_2$ column densities.  However, $f^N$ is most likely an underestimate of
$f$ in the more molecular regions of the cloud which contain higher concentrations of \hhh.  This is because atomic 
hydrogen is more widely distributed than molecular
hydrogen and column densities measure material along the entire line of sight \citep{sno06}.  
Since the measurement of $N$(H) includes
material not associated with H$_2$, $f^N$ underestimates the H$_2$ fraction in the molecular region.  For
sightlines where measurements are lacking we use $f=0.67$, the value for which the column densities of
H and H$_2$ are equal.  

When possible, estimates of the number density of hydrogen nuclei, $n_{\rm H}$, are adopted from the literature based on various
atomic and molecular diagnostics.  In eleven of our sightlines 
\citet{son07} used the observed rotational excitation of C$_2$ to infer the sum of the H and H$_2$ number densities.  This
was done by comparing models with various temperatures and number densities to the measured column densities of all the
excited states and choosing the best fit.  For sightlines where the average value of $f$ is known, they converted 
$n({\rm H}+{\rm H}_2)$ to $n_{\rm H}$.  For sightlines where $f$ is not known, our adopted value of $f=0.67$
is used to perform the conversion.  In two additional cases \citet{jur75} measured column densities of H and the $J=4$ 
excited level of H$_2$,
and with some assumptions estimated the product $Rn_{\rm H}$, where $R$ is the rate at which H$_2$ forms on grains.  Adopting
a typical value for $R$ then allowed for the computation of $n_{\rm H}$.  In one more sightline (40 Per)
\citet{jen83} estimated the 
thermal pressure from measurements of the $J=0$, 1, and 2 fine-structure levels of C for a kinetic
temperature of 80~K.  
Using this pressure estimate and the H$_2$ temperature ($T_{01}=63$~K), we calculate $n_{\rm H}$.  
Unfortunately, the results obtained for a given sightline by using each of these methods can be
significantly different.  For example, in the sightline toward $\zeta$ Oph \citet{son07}, \citet{jur75}, and \citet{jen83}
derived values of 215~cm$^{-3}$, 90~cm$^{-3}$, and 117~cm$^{-3}$, respectively for $n_{\rm H}$.  Because of the uncertainties
involved with each method and the different final results, the number densities we use are probably uncertain by about a factor
of two.  For cases where  no number density has been determined, a value of 250 cm$^{-3}$ is adopted.  

Again assuming a uniform distribution of gas in each cloud, we divide the total hydrogen column 
densities by these number densities to obtain pathlengths:
\begin{equation}
L = \frac{N_{\rm H}}{n_{\rm H}} = \frac{N({\rm H})+2N({\rm H_2})}{n_{\rm H}}.
\label{eq9}
\end{equation}
In sightlines where no H and H$_2$ column densities have been determined, the total hydrogen column density is estimated
from the color excess and the relation
$N_H\approx E(B-V)\times5.8\times10^{21}$~cm$^{-2}$ \citep{boh78,rac02}.  As the pathlength is calculated directly from the
hydrogen number density, it is also uncertain by about a factor of two.  
Using the above relations and approximations, we calculate the ionization rate of molecular hydrogen, 
$\zeta_2$.  However, most studies examine the primary cosmic-ray ionization rate per hydrogen atom,
$\zeta_p$.  While the ionization efficiencies of H and H$_2$ are dependent on factors such as 
the helium abundance and
ratio of molecular to atomic hydrogen \citep{dal99}, we adopt a more simplified approach and use the
conversion factor given by \citet{gla74}:
\begin{equation}
\zeta_2 = 2.3\zeta_p.
\label{eq10}
\end{equation}
This conversion stems from the fact that H$_2$ contains two hydrogen atoms, so the ionization rate
is nearly twice as high.  Also, $\zeta_p$ only accounts for the initial (primary) cosmic-ray ionization
while $\zeta_2$ includes ionization from energetic secondary electrons which were created in the first
ionization event.  With equation (\ref{eq10}) we convert our values of $\zeta_2$ to $\zeta_p$ so that 
they can be directly compared to previous observations.  The resulting values for the primary
cosmic-ray ionization rate and the specific estimates used for each sightline are shown in
Table \ref{tbl4}.  For completeness, we have performed the same analysis for ten sightlines from
\citet{mcc02} and also included the results in Table \ref{tbl4}.

With all of the assumptions we have made, it is important to investigate the uncertainties that will
propagate to the cosmic-ray ionization rate.  Substituting equations (\ref{eq9}) and (\ref{eq10}) into
equation (\ref{eq7}) gives the primary cosmic-ray ionization rate as

\begin{equation}
\zeta_p = \frac{2}{2.3}N({\rm H}_3^+)\frac{n_{\rm H}}{f}\frac{k_e}{N_{\rm H}}\left[\frac{n(e)}{n_{\rm H}}\right].
\label{eq11}
\end{equation}

In this equation, the molecular hydrogen fraction, $f$, and the number density of hydrogen nuclei, $n_{\rm H}$,
are the two most uncertain parameters (note that the $n_{\rm H}$ in the denominator is part of the 
ratio $n(e)/n_{\rm H}$ which is well determined).
Because $\zeta_p$ is directly
proportional to $n_{\rm H}$, any increase or decrease in $n_{\rm H}$ produces a corresponding increase or
decrease in $\zeta_p$.  As previously mentioned, the uncertainty in $n_{\rm H}$ is probably about a factor of 2.  
On the other hand, $\zeta_p$ is inversely related to $f$.  The H$_2$ fraction
is by definition between zero and one, and most of our measured values are around 0.5.  For sightlines with
the adopted value of 0.67, the maximum increase is a factor of 1.5.  We take this factor to be an
approximation for the uncertainty in $f$.  Because $f^N$ should always be an underestimate of $f$ in measured
sightlines, we only consider increasing the molecular hydrogen fraction for those sightlines.  
Taking into account the uncertainties in both $n_{\rm H}$ and $f$, the true value of $\zeta_p$ in sightlines
with measurements of $f^N$ is likely between one third and twice our derived estimate of $\zeta_p$.
For sightlines with no measured H$_2$ fraction, we allow $f$ to
vary both up or down by a factor of 1.5.  This results in a possible cosmic-ray ionization rate
between  $\zeta_p$/3 and 3$\zeta_p$.  These limits arise when the most extreme variations in both
$f$ and $n_{\rm H}$ are substituted into equation (\ref{eq11}).  However, $f$ tends to be higher when $n_{\rm H}$ is higher 
because the rate of H$_2$ formation
scales as the square of $n_{\rm H}$.
This suggests that it is probable that $f$ and $n_{\rm H}$ will vary in the same way, so that the 
above analysis most likely overestimates the range of possible ionization rates.

\subsection{\hhh\  Temperature}
Another property of the gas to be examined is the excitation temperature, determined from relative populations of the
different rotational states of \hhh.  For the two lines we have observed the temperature may be determined from the equation
\begin{equation}
\frac{N_{ortho}}{N_{para}} = \frac{g_{ortho}}{g_{para}}e^{-\Delta E/kT} = 2e^{-32.87/T}
\label{eq12}
\end{equation}
taken from \citet{mcc98}.  In this case {\it ortho} refers to the population of the (1,0) state and {\it para} to the 
(1,1) state.  The $g$'s are statistical weights, $\Delta E$ is the energy difference between the
states, $k$ is the Boltzmann constant, and $T$ is the temperature.  Excitation temperatures derived from this
equation are shown in Table \ref{tbl4}.

If the rotational (de-)excitation of \hhh\ is dominated by collisions with H$_2$, then the temperature
measurements from both species should be similar.  However, this is not the case.  Most H$_2$ temperatures are around 60~K
while the \hhh\ temperatures are typically about 30~K.  This same discrepancy was described by \citet{mcc03}.  In their model
calculation of \hhh\ thermalization, \citet{oka04} have shown that the (1,1)/(1,0) excitation temperature is always lower than
the H$_2$ temperature because of cooling by fast spontaneous emission from the (2,2) to (1,1) state.  For the typical cloud
conditions in this paper ($n_{\rm H}\sim250$~cm$^{-3}$, $T\sim60$~K) the model of \citet{oka04} produces an \hhh\ excitation
temperature of about 50-55~K which is significantly higher than the observed values of about 30~K.  
The source of this discrepancy remains unclear.

\section{Discussion}
\subsection{Inferred Ionization Rates}

Values of the cosmic-ray ionization rate for diffuse clouds observed here as well as those determined for other diffuse
clouds observed previously by us are given in the right hand column of Table \ref{tbl4}. The detected values in the lines of
sight to fourteen sources cover the range 0.5--3.2$\times10^{16}$~s$^{-1}$. Upper limits, which are given for fifteen
diffuse clouds, are consistent with this range of ionization rates, with the possible exception of HD~168607.  While most of the
detections of \hhh\ are confined to the Galactic plane, they are widely dispersed in Galactic longitude.  We therefore 
conclude that the values of the cosmic-ray ionization
rate listed in Table \ref{tbl4} are typical for Galactic diffuse interstellar clouds.

A few of the sightlines we investigated have been studied previously to derive cosmic-ray ionization rates.  All of these
studies used column densities of either OH, HD, or both in their calculations.  Because the formation pathways of OH and HD
include the ionization of atomic hydrogen, they can be used to determine the H ionization rate. Most of these studies
\citep{bla77,bla78,har78a,fed96} then derived the primary cosmic-ray ionization rate from the H ionization rate, but
\citet{odo74} did not because they still considered ionization via X-rays to be important.  Our values of the primary
ionization rate
for $\zeta$~Per, $o$~Per, $\epsilon$~Per, $\xi$~Per, and $\zeta$~Oph are shown in 
Table \ref{tbl5} along with the rates derived from OH and HD measurements as well as 
cloud modeling.  For $\zeta$~Per our value is over an order of magnitude larger than
those reported by \citet{har78b} and \citet{fed96}.  While the rest of our new measurements in Table \ref{tbl5} 
are only upper limits, these  are also typically orders of magnitude larger than previously
published values.  The only exception is $o$~Per where both papers cite
values of $\zeta_p$ about one fourth to one half our upper limit.

Various model calculations were performed by \citet{van86} to investigate three of the sightlines that we
study here: $\zeta$~Oph, $\zeta$~Per, and $o$~Per.  In creating these models they used the most recent measurements of rate
constants and the column densities of diagnostic species such as H and H$_2$ as input parameters.  By varying a few uncertain
parameters, they would then generate lists of predicted column densities for many atomic and molecular species under slightly
different conditions.  When their paper was written, it was believed that the \hhh-electron recombination 
rate constant was much lower than the currently accepted value.  
\citet{smi84} reported an upper limit correspnding to $10^{-7}$~cm$^3$~s$^{-1}$ at $T=40$~K, and \citet{ada87} lowered
the upper limit to $10^{-11}$~cm$^3$~s$^{-1}$ at $T=80$~K.
Due to the wide range of possible recombination rate constants, \citet{van86} performed
calculations using both $10^{-7}$ and $10^{-10}$~cm$^3$~s$^{-1}$.  The cosmic-ray ionization rates from their paper 
listed in Table \ref{tbl5} were computed by
determining $\zeta_p$ necessary to reproduce observed OH column densities when $k_e=10^{-7}$ cm$^3$ s$^{-1}$.  We choose to
compare these ionization rates to ours because we obtain $k_e=1.6\times10^{-7}$ cm$^3$ s$^{-1}$ when $T=60$ K
is used as the input temperature in equation (\ref{eq8}).
The value of $\zeta_p$ inferred by \citet{van86} is about the same as ours for $\zeta$~Per, but the lower 
limits they derived for $\zeta$~Oph and $o$ Per are larger than our upper limits for both of those sightlines.

For their models that used $k_e=10^{-10}$ cm$^3$ s$^{-1}$, \citet{van86} obtained cosmic-ray ionization rates
that are about a factor of 1 to 5 times smaller than ours.  From these models they also predicted 
the column density of \hhh\ along each sightline.  Their results are all on the order of
$N$(\hhh)~$\sim10^{14}$ cm$^{-2}$, which is a few times larger than the observed column densities or upper
limits in any of these sightlines.  
Because \citet{van86} use only a slightly smaller cosmic-ray ionization rate (corresponding to the formation
rate) but a much smaller recombination rate (corresponding to the destruction rate), their prediction of an 
\hhh\ column density similar to observed values seems somewhat serendipitous.  In addition to the overestimate
of the \hhh\ column density, a small \hhh-electron recombination rate constant may have further consequences.  
\citet{dal06} noted that a small value of $k_e$ may have been responsible for underestimates of the primary 
cosmic-ray ionization rate in the past.  This is because a slower destruction rate requires a slower formation
rate to produce a given abundance.

In addition to the slow recombination rate, there are some other possible explanations for differences 
between the cosmic-ray ionization rate inferred from \hhh\ and those inferred from OH and HD.  
\citet{lep04} pointed out that the rate constant associated
with the endothermic charge transfer from H$^+$ to O varies over the temperatures typically
associated with diffuse clouds.  This means that the OH production rate is temperature dependent.  
The ionization rates towards $\zeta$ Per and $\zeta$ Oph quoted in \citet{har78b} were derived using 
temperatures of 120 K and 110 K, respectively, for the warm components of the cloud models along each
sightline \citep{bla78,bla77}.  As these temperatures are about twice as
large as the values determined from H$_2$, their OH production is much more efficient.  The
result is a smaller cosmic-ray ionization rate needed to produce the observed OH column density
than if a lower temperature had been used.  This problem was addressed by the later models of 
\citet{van86} where tempertaure and density were varied as functions of cloud depth.

\citet{lep04} went on to make a comprehensive chemical model of the cloud towards
$\zeta$~Per.  They determined the value of
$\zeta_p$ that would best reproduce all observed atomic and molecular column densities to be 
$2.5\times 10^{-16}$ s$^{-1}$, which is in good agreement with our estimate
of $3.2\times 10^{-16}$ s$^{-1}$.  The difference in these values may arise
because we assume a uniform distribution of gas while \citet{lep04} invoke a three phase  
model which includes diffuse gas, dense gas, and magnetohydrodynamic shocks.

Two different effects may lead to underestimates of $\zeta_p$ from measurements of HD.  The first has
to do with an overestimate of the total deuterium to hydrogen ratio $n_{\rm D}$/$n_{\rm H}$.  This ratio can be
used to estimate the molecular deuterium fraction, $N({\rm HD})/N({\rm H}_2)$.  However,
the observed values of $N({\rm HD})/N({\rm H}_2)$ are about an order of magnitude smaller than those
predicted by $n_{\rm D}$/$n_{\rm H}$.  To explain this discrepancy, \citet{lis06} argued that the atomic deuterium  
fraction must be larger than the total deuterium fraction.  This means that approximating $n_{\rm D}$/$n_{\rm H}$ 
with  $N({\rm D})/N({\rm H})$ overestimates the 
total deuterium to hydrogen ratio.
\citet{fed96} showed that the cosmic-ray ionization rate is inversely related to the deuterium fraction,
so an overestimate of $n_{\rm D}$/$n_{\rm H}$  will underestimate $\zeta_p$.  

Secondly, \citet{lis03} emphasized the importance of grain neutralization proposed by \citet{lep88}.  
This process reduces the number of 
H$^+$ ions in the gas through charge transfer with small grains.  By lowering the abundance of H$^+$,
the production rate of HD will decrease.  This is because HD formation is dependent upon the reaction
involving the charge transfer from H$^+$ to D.  Since neutralization slows down HD production, a larger
value of $\zeta_p$ is needed to create a given abundance than if the effect were not taken into
account.  \citet{lis03} used a
model which includes grain neutralization and showed that both \hhh\ and HD column densities
can be reproduced with a single ionization rate of $\zeta_p \geq 2\times 10^{-16}$ s$^{-1}$.
Since OH formation is dependent on a similar charge transfer reaction,
grain neutralization and thus a larger cosmic-ray ionization rate may be necessary in its analysis as
well.

\citet{mcc03} studied \hhh\ in the sightline towards $\zeta$~Per.  Using nearly the same
analysis as this paper, they inferred a value of $\zeta_2 = 1.2 \times 10^{-15}$~s$^{-1}$ 
which is equivalent to $\zeta_p = 5.2\times 10^{-16}$~s$^{-1}$ shown in Table \ref{tbl5}.  This higher 
ionization rate is due to
various differences in input parameters.  In terms of the parameters in this paper, 
\citet{mcc03} used 1.5$k_e$, 1.2$N$(\hhh), 1.2$n_{\rm H}$, and 0.8$n(e)/n_{\rm H}$ for the 
following reasons.  The \hhh-electron
recombination rate constant differs because they approximated the electron temperature with the 
\hhh\ temperature instead of the H$_2$ temperature in equation (\ref{eq8}).  Further observations have
more than doubled the total integration time so that the spectrum and \hhh\ column density change
slightly between papers.  The value of $n_{\rm H}$ used by \citet{mcc03} was an average
number density computed from various measurements, whereas the value used in this 
paper comes from the C$_2$ analysis of \citet{son07}.  Finally, we have adopted a
single value of $n(e)/n_{\rm H}$ to be used in all calculations while they used H$_2$ and C$^+$ 
column densities measured towards $\zeta$~Per.  

While all of the observations and models above are viable methods for finding the cosmic-ray
ionization rate, we believe that the use of \hhh\ should produce the best results due to
its relatively simple chemistry.  Using either OH or HD to calculate $\zeta_p$ requires more 
measurements, more assumptions, and more variable parameters than using \hhh.  More parameters
give the opportunity for a greater uncertainty to accumulate during the calculation.  Fewer uncertainties
coupled with advances in instrumentation lead us to speculate that the cosmic-ray ionization rates
inferred from \hhh\ may be the most accurate to date for diffuse clouds.  However, improved estimates of
$f$ and $n_{\rm H}$, the two most uncertain values in our calculations, would make \hhh\ an even better 
probe of the cosmic-ray ionization rate.  

\subsection{Theoretical Ionization Rates}

Several theoretical calculations of $\zeta_p$ have been performed in the last half-century
\citep{hay61,spi68,nat94,web98}.  In these papers the authors derived a cosmic-ray ionization
rate starting from the observed flux of cosmic-rays in our solar system.  Unfortunately, there
are large uncertainties associated with this method.  The cosmic-ray spectrum is well measured
above about 1~GeV, but lower energy particles are deflected from the inner solar system by the
magnetic field coupled to the solar wind.  The particles which are most important for ionizing
species in diffuse clouds are likely those with energies from about 2 to 10~MeV.  Since this
portion of the spectrum cannot be directly measured, the flux at low energies must somehow be
extrapolated from existing data.  \citet{hay61} assumed that the power law which applies to
the flux of high energy cosmic-rays continues down to 10 MeV where the spectrum peaks and then
decreases linearly with energy.  From these assumptions they derived an ionization rate of
$10^{-15}$~s$^{-1}$.  \citet{spi68}, however, fit a curve to measurements of 
cosmic-rays with energies near 100~MeV that also matches the high energy spectrum power law.  
With this method, their
spectrum peaks around 100 MeV and falls off for lower energies.  The result of using their fit 
is a lower limit of $6.8\times10^{-18}$~s$^{-1}$.  In the same paper they derived
an upper limit of $1.2\times10^{-15}$~s$^{-1}$ via arguments that low energy
cosmic-ray protons are accelerated in Type~I supernova shells.  \citet{web98} used data from the
{\it Pioneer} and {\it Voyager} spacecraft as they travelled outward in the solar system where the
weaker solar wind allows for the detection of lower-energy cosmic-rays.  These data were then
combined with previous observations to infer the interstellar proton spectrum.  Using this proton
spectrum and a heavy nuclei spectrum both with low energy cut-offs at 10~MeV and an electron spectrum
cut-off below 2~MeV, \citet{web98} calculated the primary cosmic-ray ionization rate to be $(3-4)\times
10^{-17}$ s$^{-1}$.  Our ionization rates fall
neatly within the bounds formed by these studies and so are not inconsistent with constraints based on direct cosmic-ray
measurements and theoretical particle physics.

\subsection{The Ionization Rate in Dense Clouds}

In contrast to our findings in diffuse clouds, the cosmic-ray ionization rate in dense clouds does
seem to agree with the canonical value.  Observations of \hhh\ towards dense clouds have found column
densities roughly the same as those seen in diffuse clouds \citep{geb96,mcc99}.  These measurements    
have been used to calculate the product $\zeta_2L$.  When $\zeta_2$ is taken to be the canonical value
of $\sim 3\times10^{-17}$~s$^{-1}$, the resulting pathlength is on the order of a parsec.  This is a
typical size for dense clouds as measured by other methods such as extinction mapping.  Since \hhh\
should be a reliable tracer for the cosmic-ray ionization rate in both environments, there must be some
mechanism causing the difference between dense and diffuse clouds.  
One possibility examined by both \citet{ski76} and \citet{pad05}
is cosmic-ray self-confinement.  In this process cosmic-rays generate 
Alfv\'{e}n waves which can effectively confine the lower energy particles ($\lesssim100$~MeV) to diffuse 
material, thus preventing them from entering dense clouds.  Because cosmic-rays in the 1--100~MeV range 
are the most efficient at ionization, self-confinement naturally leads to a higher ionization rate in diffuse
clouds than in dense clouds.  Another possibility is that there is a previously unrecognized high flux of 
low energy cosmic-rays that can penetrate diffuse but not dense clouds.  Assuming that typical column densities
of diffuse clouds are of order $10^{21}$~cm$^{-2}$ and those of dense clouds are of order $10^{23}$~cm$^{-2}$, 
cosmic-rays with energies $\sim$2--20 MeV \citep{cra78} would contribute to the 
ionization rate only in diffuse clouds.  As we foresee no observational techniques that would distinguish between
these two possibilities, a resolution to this question will depend on more sophisticated theoretical treatments.

\section{SUMMARY \& CONCLUSIONS}
We have surveyed twenty sightlines and detected \hhh\ along eight of them.  Column
densities are measured for these eight sightlines, and upper limits set for the remaining
twelve.  Besides a concentration near the Galactic plane, there seems to be no clear correlation 
between location in the sky and detecting \hhh, so it
is unlikely that we are observing anomalous regions in the Galaxy.  Instead, finding \hhh\ in so many 
sightlines suggests that it is ubiquitous in the diffuse ISM.

From the \hhh\ column densities and the steady state approximation, we
derive cosmic-ray ionization rates for the nineteen diffuse cloud sightlines in this study along with ten
sightlines from \citet{mcc02}.  Typical values are on the order of $\zeta_p\approx2\times10^{-16}$
s$^{-1}$, which falls within theoretical constraints.  While this is an order of magnitude 
larger than most previously inferred values, there are
several possible explanations for the discrepancy.  The most likely candidates are rate 
constants with uncertain measurements and physical or chemical effects not included in past models. 
Newer models that do take into account these factors require cosmic-ray ionization rates very similar
to our inferred values.  Coupled with these models, our widespread detection of \hhh\ in diffuse 
clouds supports the idea that the typical cosmic-ray ionization rate in such regions should be 
revised upward by about an order of magnitude.
\\

The authors would like to thank J. H. Black and H. S. Liszt for helpful comments and suggestions,
and acknowledge the staff of UKIRT and the developers of the data
reduction packages used in this paper.
The United Kingdom Infrared Telescope is operated by the Joint Astronomy Centre on behalf of the U.K. 
Particle Physics and Astronomy Research Council.
The ORAC-DR and FIGARO software packages were provided by the Starlink Project which 
is run by CCLRC on behalf of PPARC.
NI and BJM have been supported by NSF grant PHY 05-55486.  
TO has been supported by NSF grant PHY 03-54200.  TRG's research is supported by the Gemini Observatory,  
which is operated by the Association of Universities 
for Research in Astronomy, Inc., on behalf of the international Gemini partnership of Argentina, 
Australia, Brazil, Canada, Chile, the United Kingdom, and the United States of America.


\clearpage

\clearpage

\clearpage

\clearpage

\clearpage


\clearpage
\begin{deluxetable}{ccccc}
\tablecaption{Observations \label{tbl1}}
\tablehead{
 & & \colhead{Date(s) of} &
 & \colhead{Integration Time} \\
\colhead{Object} & \colhead{HD number} & \colhead{Observation} &
\colhead{Standard} & \colhead{(sec)}
}

\startdata
$\zeta$ Oph    & 149757 & 2001 May 24 & $\beta$ Lib  &  576\\
               &        & 2001 May 25 & $\beta$ Lib  & 1344\\
HD 147889      & 147889 & 2001 May 24 & $\delta$ Sco & 5760\\
$\lambda$ Cep  & 210839 & 2001 May 24 & $\alpha$ Lyr & 4320\\
HD 169454      & 169454 & 2001 May 25 & $\beta$ Lib  & 4224\\
W40 IRS 1a     &    ... & 2001 May 25 & $\eta$ Oph   & 1800\\
               &        & 2001 May 26 & $\delta$ Sco & 1800\\
$o$ Sco        & 147084 & 2001 May 26 & $\alpha$ Lyr &  960\\
HD 168625      & 168625 & 2001 May 26 & $\eta$ Oph   & 1440\\
HD 229059      & 229059 & 2001 May 27 & $\alpha$ Cyg & 1800\\
               &        & 2001 Sept 5 & $\alpha$ Cyg & 1800\\
BD -14 5037    &    ... & 2001 May 28 & $\eta$ Oph   & 2016\\
HD 20041       &  20041 & 2001 Sept 5 & $\beta$ Per  & 1728\\
HD 21389       &  21389 & 2001 Sept 5 & $\beta$ Per  & 1536\\
               &        & 2002 Dec 30 & $\delta$ Per & 1152\\
               &        & 2002 Dec 31 & $\beta$ Per  & 1536\\
               &        & 2003 Jan 1  & $\eta$ Tau   & 1152\\
$\zeta$ Per    &  24398 & 2001 Sept 5 & $\beta$ Per  & 2304\\
               &        & 2002 Dec 30 & $\beta$ Per  & 1152\\
               &        & 2002 Dec 31 & $\beta$ Per  & 1152\\
               &        & 2003 Jan 1  & $\delta$ Per & 1440\\
$o$ Per        &  23180 & 2002 Dec 30 & $\beta$ Per  & 2304\\
               &        & 2002 Dec 31 & $\beta$ Per  & 1152\\
$\xi$ Per      &  24912 & 2002 Dec 30 & $\beta$ Per  & 1920\\
               &        & 2002 Dec 31 & $\beta$ Per  & 1920\\
               &        & 2003 Jan 1  & $\delta$ Per & 1536\\
X Per          &  24534 & 2002 Dec 31 & $\beta$ Per  & 1920\\
               &        & 2004 Jan 22 & HD 17573     & 3600\\
               &        & 2005 Jan 5  & HD 17573     & 4320\\
               &        & 2005 Jan 6  & $\eta$ Aur   & 2880\\
               &        & 2005 Mar 3  & $\eta$ Aur   & 5040\\
               &        & 2005 Mar 4  & $\eta$ Aur   & 5040\\
62 Tau         &  27778 & 2003 Jan 1  & $\beta$ Per  & 2688\\
$\epsilon$ Per &  24760 & 2004 Jan 23 & $\eta$ Aur   & 1800\\
               &        & 2005 Jan 5  & $\eta$ Aur   & 1440\\
               &        & 2005 Jan 6  & $\eta$ Aur   & 1440\\
40 Per         &  22951 & 2004 Jan 23 & HD 17573     & 3600\\
               &        & 2005 Jan 6  & $\eta$ Aur   & 2880\\
Red Rectangle  &  44179 & 2005 Jan 6  & $\kappa$ Ori &  720\\
HD 21483       &  21483 & 2005 Jan 25 & HD 17573     & 5760\\ 
			     
\enddata		     
\end{deluxetable}

\clearpage
\begin{deluxetable}{lrcc}
\tablecaption{Previous Measurements of ISM Gas Velocities \label{tbl2}}
\tablehead{ & \colhead{$v_{\rm LSR}$} & & \\
\colhead{Object} & \colhead{(km s$^{-1}$)} & \colhead{Species} & \colhead{Reference}}

\startdata
HD 20041       & -1.6 & K \textsc{i} & 1 \\
HD 21389       & -0.5 & CH & 1 \\
$\zeta$ Per    &  6.9 & CH & 1 \\
X Per          &  6.6 & CH & 1 \\
HD 169454      &  5.3 & CH & 1 \\
               &   90 & Na \textsc{i} \& Ca \textsc{ii} & 2 \\
HD 229059      & 4.04 & K \textsc{i} & 3 \\
BD -14 5037    &  8.2 & C$_2$ & 4 \\
               & 18.2 & C$_2$ & 4 \\
W40 IRS 1a     &    8 & $^{13}$CO\tablenotemark{\dag} & 5 \\
HD 21483       & 10.3 & CH & 1 \\
40 Per         &  6.8 & K \textsc{i}& 3 \\
$o$ Per        &  7.3 & CH & 1 \\
$\epsilon$ Per &  2.2 & K \textsc{i}& 6 \\
$\xi$ Per      & 0.55 & K \textsc{i} & 6 \\
62 Tau         &  4.8 & CH & 1 \\
Red Rectangle  & -0.8 & K \textsc{i}\tablenotemark{\dag} \& O \textsc{i}\tablenotemark{\dag}& 7 \\
$o$ Sco        & 2.29 & K \textsc{i} & 6 \\
HD 147889      &  2.7 & K \textsc{i} & 1 \\
$\zeta$ Oph    & -1.0 & CH & 1 \\
HD 168625      &    6 & Ca \textsc{ii} & 8 \\
$\lambda$ Cep  & -1.7 & CH & 1 \\
\enddata
\tablerefs{(1) Welty (private communication); (2) \citet{fed92}; (3) \citet{cha82}; (4) \citet{gre86};
(5) \citet{cru82}; (6) \citet{wel01}; (7) \citet{hob04}; (8) \citet{ric74}}
\tablecomments{$v_{\rm LSR}$ is the velocity of the interstellar gas in the local standard of rest frame.  Unless noted,
all lines were measured in absorption.}
\tablenotetext{\dag}{measured in emission}
\end{deluxetable}

\clearpage
\begin{deluxetable}{llcccccc}
\tablecaption{Absorption Line Parameters \label{tbl3}}
\tablehead{
 & & \colhead{$v_{\rm LSR}$} & \colhead{FWHM} & \colhead{$W_{\lambda}$} & \colhead{$\sigma(W_{\lambda})$} &
\colhead{$N$(H$_3^+)$} & \colhead{$\sigma(N)$} \\
\colhead{Object} & \colhead{Transition} & \colhead{(km s$^{-1}$)} & \colhead{(km s$^{-1}$)} &
\colhead{(\AA)} & \colhead{(\AA)} & \colhead{($10^{14}$ cm$^{-2}$)} & \colhead{($10^{14}$ cm$^{-2}$)}
}

\startdata
HD 20041       & $R(1,1)^u$   &  -1.4 &  8.4 & 0.017    & 0.004 & 0.70    & 0.15  \\
               & $R(1,0)$     &  -1.5 &  9.6 & 0.036    & 0.004 & 0.91    & 0.10  \\
HD 21389       & $R(1,1)^u$   &  -1.9 & 11.9 & 0.009    & 0.002 & 0.39    & 0.07  \\
               & $R(1,0)$     &  -0.2 & 15.6 & 0.016    & 0.002 & 0.41    & 0.05  \\
$\zeta$ Per    & $R(1,1)^u$   &   7.5 &  9.7 & 0.010    & 0.001 & 0.43    & 0.05  \\
               & $R(1,0)$     &   6.8 &  8.8 & 0.010    & 0.001 & 0.26    & 0.03  \\
X Per          & $R(1,1)^u$   &   8.3 & 11.6 & 0.011    & 0.002 & 0.46    & 0.10  \\
               & $R(1,0)$     &   5.7 &  9.6 & 0.012    & 0.002 & 0.31    & 0.06  \\
HD 169454      & $R(1,1)^u$   &   2.6 & 11.1 & 0.005    & 0.002 & 0.21    & 0.08  \\
               & $R(1,0)$     &   2.1 & 10.8 & 0.014    & 0.002 & 0.35    & 0.05  \\
HD 229059      & $R(1,1)^u$   &   5.9 & 12.2 & 0.058    & 0.003 & 2.42    & 0.14  \\
               & $R(1,0)$     &   4.4 & 12.4 & 0.059    & 0.003 & 1.48    & 0.09  \\
BD -14 5037    & $R(1,1)^u$   &  17.6 &  9.6 & 0.010    & 0.003 & 0.40    & 0.13  \\
               & $R(1,0)$     &  18.6 & 10.4 & 0.009    & 0.003 & 0.23    & 0.08  \\
W40 IRS 1a     & $R(1,1)^u$   &   7.5 & 10.8 & 0.051    & 0.008 & 2.12    & 0.33  \\
               & $R(1,0)$     &   8.2 &  6.6 & 0.050    & 0.006 & 1.26    & 0.16  \\
HD 21483       & $R(1,1)^u$   &   ... &   10 & $<$0.036 & ...   & $<$1.53 & ...   \\
               & $R(1,0)$     &   ... &   10 & $<$0.036 & ...   & $<$0.93 & ...   \\
40 Per         & $R(1,1)^u$   &   ... &   10 & $<$0.015 & ...   & $<$0.60 & ...   \\
               & $R(1,0)$     &   ... &   10 & $<$0.015 & ...   & $<$0.36 & ...   \\
$o$ Per        & $R(1,1)^u$   &   ... &   10 & $<$0.009 & ...   & $<$0.42 & ...   \\
               & $R(1,0)$     &   ... &   10 & $<$0.009 & ...   & $<$0.27 & ...   \\
$\epsilon$ Per & $R(1,1)^u$   &   ... &   10 & $<$0.009 & ...   & $<$0.39 & ...   \\
               & $R(1,0)$     &   ... &   10 & $<$0.009 & ...   & $<$0.24 & ...   \\
$\xi$ Per      & $R(1,1)^u$   &   ... &   10 & $<$0.009 & ...   & $<$0.33 & ...   \\
               & $R(1,0)$     &   ... &   10 & $<$0.009 & ...   & $<$0.21 & ...   \\
62 Tau         & $R(1,1)^u$   &   ... &   10 & $<$0.045 & ...   & $<$1.89 & ...   \\
               & $R(1,0)$     &   ... &   10 & $<$0.045 & ...   & $<$1.14 & ...   \\
Red Rectangle  & $R(1,1)^u$   &   ... &   10 & $<$0.006 & ...   & $<$0.21 & ...   \\
               & $R(1,0)$     &   ... &   10 & $<$0.006 & ...   & $<$0.15 & ...   \\
$o$ Sco        & $R(1,1)^u$   &   ... &   10 & $<$0.009 &  ...  & $<$0.36 & ...   \\
               & $R(1,0)$     &   ... &   10 & $<$0.009 &  ...  & $<$0.21 & ...   \\
HD 147889      & $R(1,1)^u$   &   ... &   10 & $<$0.009 &  ...  & $<$0.39 & ...   \\
               & $R(1,0)$     &   ... &   10 & $<$0.009 &  ...  & $<$0.24 & ...   \\
$\zeta$ Oph    & $R(1,1)^u$   &   ... &   10 & $<$0.003 &  ...  & $<$0.18 & ...   \\
               & $R(1,0)$     &   ... &   10 & $<$0.003 &  ...  & $<$0.12 & ...   \\
HD 168625      & $R(1,1)^u$   &   ... &   10 & $<$0.012 &  ...  & $<$0.54 & ...   \\
               & $R(1,0)$     &   ... &   10 & $<$0.012 &  ...  & $<$0.33 & ...   \\
$\lambda$ Cep  & $R(1,1)^u$   &   ... &   10 & $<$0.012 &  ...  & $<$0.54 & ...   \\
               & $R(1,0)$     &   ... &   10 & $<$0.012 &  ...  & $<$0.33 & ...   \\
\enddata
\tablecomments{$v_{\rm LSR}$ is the observed line of sight velocity in the local standard of rest
frame.  FWHM is the line full width at half maximum (for the purpose of
calculating column density upper limits, the FWHM is assumed to be 10 km s$^{-1}$ for all
spectra without absorption lines).  $W_{\lambda}$ is the equivalent width of the line in \AA ngstr\"{o}ms.
$\sigma(W_{\lambda})$ is the one standard deviation uncertainty of the equivalent width.
Upper limits for the equivalent width were found by taking 3$\times\sigma(W_{\lambda})$.
$N$(H$_3^+)$ is the \hhh\ column density. $\sigma(N)$ is the one standard deviation uncertainty 
of the \hhh\ column density.  Upper limits for the column density were found by taking 3$\times\sigma(N)$.}

\end{deluxetable}

\clearpage

\begin{deluxetable}{lccccccccc}
\rotate
\tabletypesize{\footnotesize}
\tablecaption{Sightline Parameters \label{tbl4}}
\tablehead{
 & \colhead{$N$(H$_3^+)_{tot}$} & \colhead{$\sigma(N)$} & \colhead{$T$(\hhh)} & \colhead{$E(B-V)$} & 
\colhead{$N_{\rm H}$} & & \colhead{$n_{\rm H}$} & \colhead{$L$} & \colhead{$\zeta_p$} \\
\colhead{Object} & \colhead{($10^{14}$ cm$^{-2}$)} & 
\colhead{($10^{14}$ cm$^{-2}$)} & \colhead{(K)} & \colhead{(mag)} & \colhead{($10^{21}$ cm$^{-2}$)} & 
\colhead{$f$} & \colhead{(cm$^{-3}$)} &
\colhead{(pc)} & \colhead{($10^{-16}$ s$^{-1}$)}
}

\startdata
HD 20041       & 1.6                  & 0.18                  & 76\tablenotemark{c} & 0.72\tablenotemark{d} & 
4.18\tablenotemark{m} & 0.67\tablenotemark{q} & 250\tablenotemark{t} & 5.4 & 2.9 \\

HD 21389       & 1.0                  & 0.08                  & 51 & 0.57\tablenotemark{d} & 
3.31\tablenotemark{m} & 0.67\tablenotemark{q} & 250\tablenotemark{t} & 4.3 & 1.8 \\

$\zeta$ Per    & 0.7                  & 0.06                  & 28 & 0.31\tablenotemark{d} & 
1.59\tablenotemark{n}\tablenotemark{o} & 0.60\tablenotemark{r} & 215\tablenotemark{u} & 2.4 & 3.2 \\

X Per          & 0.8                  & 0.17                  & 30 & 0.59\tablenotemark{d} & 
2.20\tablenotemark{n}\tablenotemark{p} & 0.76\tablenotemark{r} & 325\tablenotemark{u} & 2.2 & 3.1 \\

HD 169454      & 0.6                  & 0.09                  & 180\tablenotemark{c} & 1.12\tablenotemark{d} & 
6.50\tablenotemark{m} & 0.50\tablenotemark{s} & 265\tablenotemark{u} & 7.9 & 0.9 \\

HD 229059      & 3.9                  & 0.16                  & 28 & 1.71\tablenotemark{d} & 
9.92\tablenotemark{m} & 0.67\tablenotemark{q} & 250\tablenotemark{t} & 13  & 2.9 \\

BD -14 5037    & 0.6                  & 0.16                  & 26 & 1.55\tablenotemark{e} & 
8.99\tablenotemark{m} & 0.67\tablenotemark{q} & 250\tablenotemark{t} & 12  & 0.5 \\

W40 IRS 1a     & 3.4                  & 0.37                  & 27 & 2.90\tablenotemark{f} & 
16.8\tablenotemark{m} & 0.67\tablenotemark{q} & 250\tablenotemark{t} & 22  & 1.5 \\

WR 104         & 2.3\tablenotemark{a} & 0.25\tablenotemark{a} & 38\tablenotemark{a} & 2.10\tablenotemark{g} & 
12.2\tablenotemark{m} & 0.67\tablenotemark{q} & 250\tablenotemark{t} & 16  & 1.4 \\

WR 118         & 6.5\tablenotemark{a} & 0.18\tablenotemark{a} & 40\tablenotemark{a} & 4.13\tablenotemark{g} & 
24.0\tablenotemark{m} & 0.67\tablenotemark{q} & 250\tablenotemark{t} & 31  & 2.0 \\

WR 121         & 2.2\tablenotemark{a} & 0.28\tablenotemark{a} & ...& 1.68\tablenotemark{g} & 
9.74\tablenotemark{m} & 0.67\tablenotemark{q} & 250\tablenotemark{t} & 13  & 1.7 \\

Cyg OB2 12     & 3.8\tablenotemark{a} & 0.36\tablenotemark{b} & 27\tablenotemark{a} & 3.35\tablenotemark{h} & 
19.4\tablenotemark{m} & 0.67\tablenotemark{q} & 300\tablenotemark{v} & 21  & 1.8 \\

Cyg OB2 5      & 2.6\tablenotemark{a} & 0.19\tablenotemark{a} & 47\tablenotemark{a} & 1.99\tablenotemark{h} & 
11.5\tablenotemark{m} & 0.67\tablenotemark{q} & 225\tablenotemark{v} & 17  & 1.5 \\

HD 183143      & 2.3\tablenotemark{a} & 0.08\tablenotemark{a} & 31\tablenotemark{a} & 1.28\tablenotemark{i} & 
7.42\tablenotemark{m} & 0.67\tablenotemark{q} & 250\tablenotemark{t} & 9.6 & 2.3 \\


HD 21483       & $<$2.2                & ...                  & ... & 0.56\tablenotemark{d} & 
3.25\tablenotemark{m} & 0.67\tablenotemark{q} & 250\tablenotemark{t} & 4.2 & $<$5.7 \\

40 Per         & $<$0.9                & ...                  & ... & 0.24\tablenotemark{j} & 
1.67\tablenotemark{n}\tablenotemark{o} & 0.35\tablenotemark{r}  & 80\tablenotemark{w} & 6.7 & $<$2.6 \\

$o$ Per        & $<$0.6                & ...                  & ... & 0.31\tablenotemark{d} & 
1.52\tablenotemark{n}\tablenotemark{o} & 0.54\tablenotemark{r} & 265\tablenotemark{u} & 1.9 & $<$5.0 \\

$\epsilon$ Per & $<$0.5                & ...                  & ... & 0.09\tablenotemark{j} & 
0.35\tablenotemark{n}\tablenotemark{o} & 0.19\tablenotemark{r}  & 15\tablenotemark{x} & 7.5 & $<$2.4 \\

$\xi$ Per      & $<$0.5                & ...                  & ... & 0.33\tablenotemark{d} & 
1.82\tablenotemark{n}\tablenotemark{o} & 0.38\tablenotemark{r} & 300\tablenotemark{x} & 2.0 & $<$4.5 \\

62 Tau         & $<$2.7                & ...                  & ... & 0.37\tablenotemark{d} & 
2.19\tablenotemark{n}\tablenotemark{p} & 0.56\tablenotemark{r} & 280\tablenotemark{u} & 2.5 & $<$14 \\


$o$ Sco        & $<$0.5                & ...                  & ... & 0.73\tablenotemark{e} & 
4.23\tablenotemark{m} & 0.67\tablenotemark{q} & 225\tablenotemark{v} & 6.1 & $<$0.9 \\

HD 147889      & $<$0.6                & ...                  & ... & 1.07\tablenotemark{d} & 
6.21\tablenotemark{m} & 0.67\tablenotemark{q} & 525\tablenotemark{v} & 3.8 & $<$1.6 \\

$\zeta$ Oph    & $<$0.3                & ...                  & ... & 0.32\tablenotemark{d} & 
1.40\tablenotemark{n}\tablenotemark{o} & 0.65\tablenotemark{r} & 215\tablenotemark{u} & 2.1 & $<$1.5 \\

HD 168625      & $<$0.8                & ...                  & ... & 1.48\tablenotemark{e} & 
8.58\tablenotemark{m} & 0.67\tablenotemark{q} & 250\tablenotemark{t} & 11  & $<$0.8 \\

$\lambda$ Cep  & $<$0.8                & ...                  & ... & 0.57\tablenotemark{d} & 
2.80\tablenotemark{n}\tablenotemark{p} & 0.49\tablenotemark{r} & 115\tablenotemark{u} & 7.8 & $<$1.3 \\

HD 168607      & $<$0.6\tablenotemark{a} & ... & ... & 1.61\tablenotemark{i} & 
9.34\tablenotemark{m} & 0.67\tablenotemark{q} & 250\tablenotemark{t} & 12  & $<$0.5 \\

HD 194279      & $<$1.2\tablenotemark{a} & ... & ... & 1.22\tablenotemark{i} & 
7.08\tablenotemark{m} & 0.67\tablenotemark{q} & 250\tablenotemark{t} & 9.1 & $<$1.3 \\

$\chi^2$ Ori   & $<$0.7\tablenotemark{a} & ... & ... & 0.44\tablenotemark{k} & 
2.55\tablenotemark{m} & 0.67\tablenotemark{q} & 250\tablenotemark{t} & 3.3 & $<$2.1 \\

P Cyg          & $<$0.6\tablenotemark{a} & ... & ... & 0.63\tablenotemark{l} & 
3.65\tablenotemark{m} & 0.67\tablenotemark{q} & 250\tablenotemark{t} & 4.7 & $<$1.2 \\
\enddata

\tablecomments{$N$(H$_3^+)_{tot}$ is total H$_3^+$ column density.  $\sigma(N)$ is one
standard deviation uncertainty of the total column density.  Upper limits for the column density were found by taking
3$\times\sigma(N)$.
$T$(\hhh) is the excitation temperature of \hhh\ as determined from the column densities of the (1,0) and (1,1) 
states.  $E(B-V)$ is the color excess.           
$N_{\rm H}$ is the column density of hydrogen nuclei.  $f$ is the molecular hydrogen fraction.
$n_{\rm H}$ is the number density of hydrogen nuclei.  $L$ is the cloud path length assuming a uniform
distribution of gas.  $\zeta_p$ is the primary cosmic-ray ionization rate.  
Upper limits on $\zeta_p$ were calculated using 3$\times\sigma(N)$.}

\tablenotetext{a}{from \citet{mcc02}}
\tablenotetext{b}{from \citet{mcc98}}
\tablenotetext{c}{these high temperatures are most likely caused by inaccurate measurements of the (1,1) state column density 
due to atmospheric interference}
\tablenotetext{d}{from \citet{tho03}}
\tablenotetext{e}{derived from method used in \citet{tho03}}
\tablenotetext{f}{derived from \citet{shu99} assuming $R_V = A_V/E(B-V) = 3.1$}
\tablenotetext{g}{derived from \citet{pen94} assuming $R_V = A_V/E(B-V) = 3.1$}
\tablenotetext{h}{from \citet{sch58}}
\tablenotetext{i}{from \citet{sno77}}
\tablenotetext{j}{from \citet{sav77}}
\tablenotetext{k}{derived from intrinsic color of \citet{weg94}}
\tablenotetext{l}{from \citet{lam83}}
\tablenotetext{m}{calculated from $N_{\rm H}\approx E(B-V)\times5.8\times10^{21}$ cm$^{-2}$ mag$^{-1}$ in \citet{boh78}}
\tablenotetext{n}{calculated from observed H and H$_2$ column densities}
\tablenotetext{o}{from B. Rachford (private communication)}
\tablenotetext{p}{from \citet{rac02}}
\tablenotetext{q}{we adopt $f = 0.67$ for sightlines without measured column densities}
\tablenotetext{r}{H$_2$ fraction derived from same sources as column densities}
\tablenotetext{s}{$f = 0.5$ assumed by \citet{son07} when calculating $n_{\rm H}$}
\tablenotetext{t}{adopted number density}
\tablenotetext{u}{from \citet{son07}}
\tablenotetext{v}{from $n$(H + H$_2$) in \citet{son07} assuming $f=0.67$}
\tablenotetext{w}{derived from pressure in \citet{jen83}}
\tablenotetext{x}{from \citet{jur75}}

\end{deluxetable}

\clearpage
\begin{deluxetable}{cccccc}
\tablecaption{Primary Cosmic-ray Ionization Rate, $\zeta_p$ ($10^{-16}$ s$^{-1}$), for Select Sightlines 
\label{tbl5}}
\tablehead{
\colhead{reference} & \colhead{$\zeta$ Per} & \colhead{$o$ Per} & \colhead{$\epsilon$ Per} &
\colhead{$\xi$ Per} & \colhead{$\zeta$ Oph}}
\startdata
1 & 3.2  & $<$5.0     & $<$2.4     & $<$4.5     & $<$1.5     \\ 
2 & 0.22 & 2.50       & 0.01       & 0.06       & 0.17       \\
3 & 0.17 & 1.30       & ...        & $\leq$0.26 & ...        \\
4 & 1-2  & $\geq$8    & ...        & ...        & $\geq$4    \\
5 & 5.2  & ...        & ...        & ...        & ...        \\
6 & 2.5  & ...        & ...        & ...        & ...        \\
\enddata
\tablecomments{The upper limits from this paper are calculated using the 3$\sigma$ uncertainty 
in the \hhh\ column density.  The value from \citet{mcc03} is found by using the conversion
factor given in equation (\ref{eq10}).}
\tablerefs{(1) this paper; (2) \citet{har78b}; (3) \citet{fed96}; (4) \citet{van86};
 (5) \citet{mcc03}; (6) \citet{lep04}}
\end{deluxetable}

\end{document}